\documentclass[12pt]{article}

\font\bbb=bbm10 scaled \magstep1
\font\fff= eufm10 scaled \magstep 1

\newtheorem{bigthm}{Theorem}   
\newtheorem{theorem}{Theorem}[section]   
\newtheorem{lemma}[theorem]{Lemma}         
\newtheorem{proposition}[theorem]{Proposition}  
\newenvironment{proof}{{\bigskip}{\sl Proof:}\quad}{\hfill{\qed}\\ \noindent}

\newenvironment{remark}{{\bigskip}{\bf Remark.\ }\rm}{\bigskip}  
\newenvironment{example}{{\bigskip}{\it Example.\ }\rm}{\bigskip}  
\newenvironment{definition}{{\bigskip}{\bf Definition\ }\rm}{\bigskip}  
\renewenvironment{abstract}{\begin{quote}{\bf Abstract.\ }\small}{\end{quote}\bigskip}
\newenvironment{acknowledgment}{\bigskip{\bf Acknowledgment.\ }\small}{\bigskip}

\def\End{{\rm End}\,}

\def\Q{\hbox{\bbb Q}}
\def\N{\hbox{\bbb N}}
\def\Mat{{\rm Mat}}

\def\tr{{\rm tr}\,}
\def\avg#1.{\langle #1 \rangle}
\def\diag#1{{\rm diag}({#1})}
\newcommand{\sym}[2]{\ensuremath{{#1}{#2}{#1}^*}}
\newcommand{\symm}[2]{\ensuremath{{#1}^*{#2}{#1}}}
\def\half{{\textstyle 1\over \textstyle 2}}

\def\rI{{\rm I}}
\def\rJ{{\rm J}}
\def\rK{{\rm K}}

\def\Z{{\cal Z }}
\def\V{{\cal V}}

\def\?#1?{\hbox{\fff #1}}

\def\vstrut{{\phantom{\biggm|}}}

\def\dg#1.{{#1}^\dagger}                                                   
\newcommand{\spow}[2]{\ensuremath{{#1}^{\vee #2}}}  
\def\hfb{\hfill\break}
\def\th{\thinspace}

\def\qed{\qquad\framebox[7pt]\medskip\noindent}

\def\eqref#1{(\ref{#1})}

\title{Zeon Algebra,  Fock Space, and Markov Chains}
\author{Ph. Feinsilver\\
Department of Mathematics\\
Southern Illinois University\\
Carbondale, IL\\
62901 USA}
\date{}

\begin{document}
\maketitle

\thispagestyle{empty}

\begin{abstract}
Fock spaces over zeons are introduced. Trace identities and a 
noncommutative ``integration-by-parts" formula are developed. 
As an application, we find a new criterion, without involving powers of the transition matrix,
 for a Markov chain to be ergodic.

\end{abstract}

\section{Introduction}
In second quantization, Fock spaces are customarily based on bosons or fermions.
Here we consider Fock spaces based on ``zeons". Zeons can be thought of as possessing a mix
of fermionic and bosonic properties: they square to zero, yet they commute. \\

In this paper we give our motivation and construction of the zeon Fock space, $\Z$.
We derive trace identities in the spirit of noncommutative integration for the degree two
component of $\Z$. Then we consider stochastic matrices specifically.
We review the results on convergence of the Markov chain corresponding to a stochastic matrix. 
Namely, the matrix being ``irreducible and aperiodic" is equivalent to ergodicity of the chain.
For stochastic matrices, the zeon noncommutative integration-by-parts formula 
leads to a new criterion for the corresponding Markov chain to be irreducible and aperiodic, hence ergodic.
Some concluding remarks round out the paper.

\section{Background}
The idea was developed from looking at semigroups of zero-one matrices representing functions
acting on a finite set. Given a set $S$ of  $n$ elements, let
$F(S)=\{f\colon S\to S\}$. For a field of scalars, take the rationals, $\Q$, as the most natural 
for our purposes and consider the vector space $V=\Q^n$ with $\End(V)$ 
the space of $n\times n $ matrices acting as endomorphisms of $V$. We have the mapping
$$ F(S)\stackrel{M}{\longrightarrow} \End(V) $$
taking $f\in F(S)$ to $M(f)\in \End(V)$ defined by

\begin{equation}\label{Xf}
(M(f))_{ij}=\cases{ 1, & if $f(i)=j$\cr 0, & otherwise}
\end{equation}

Denote by $F_\circ(S)$ the semigroup consisting of $F(S)$ with the operation of composition, 
where we compose maps to the right: for $i\in S$, $i(f_1f_2)=f_2(f_1(i))$.
The mapping $M$ gives a representation of the semigroup $F_\circ(S)$ by endomorphisms of $V$, i.e.,
$M(f_1f_2)=M(f_1)M(f_2)$.\\

Now consider the exterior algebra, $\displaystyle \bigwedge V$. 
For $W\in \End(V)$, the exterior powers of $W$, $W^{\wedge k}$, 
act on $\displaystyle V^{\wedge k}={\bigwedge}^kV$ in the standard way:

$$ W^{\wedge k}(v_1\wedge v_2\wedge\ldots \wedge v_k)=Wv_1\wedge Wv_2\wedge\ldots\wedge Wv_k\ .$$
It is clear that for general matrices, $W_1$, $W_2$, for each degree $k$,
we have a representation of $\End(V)$ by endomorphisms of $V^{\wedge k}$.
That is, for each $k$, the induced map
$$ \End(V) \to \End(V^{\wedge k})\,\colon\quad W_1\to W_1^{\wedge k},\  W_2\to W_2^{\wedge k}$$
satisfies $$(W_1W_2)^{\wedge k}=W_1^{\wedge k}\,W_2^{\wedge k}\ .$$ 
Composing with $M$, for each $k$, we have a representation of the semigroup $F_\circ(S)$ on
$V^{\wedge k}$. However, since the Grassmann
algebra has anticommuting generators, minus signs are introduced and exterior powers of $M(f)$
do not correspond to functions. \\

The remedy comes from the observation that in multiplying
exterior powers of the matrices $M(f_1)$, $M(f_2)$, there is never any cancellation effect due to
the signs, as any row has at most a single non-zero entry, $\pm 1$.  If the signs are
ignored, then we still have a homomorphism at each degree $k$, and 
products of the associated matrices will correspond to functions.\\

We conclude this section with some notation.
The details of our construction is the subject of the next section. 

\subsection{Notations}

Here we collect some notational conventions used throughout the paper.\\

We treat vectors as ``row vectors", i.e., $1\times n$ matrices, with $e_i$ as the corresponding standard
basis vectors. The column vector corresponding to $v$ is $\dg v.$.  For a matrix $M$,
$M^*$ denotes its transpose.
And $\diag{v}$ denotes the diagonal matrix with entries $v_i$ of the vector $v$.\\

The vector having components all equal to 1 is denoted $u$. And the matrix 
$J=\dg u.u$ has entries all ones. We use the convention that the identity matrix, $I$, as well as
$u$ and $J$, denote matrices of the appropriate size according to the context.\\

\section{Construction of zeon algebra}

Consider the exterior algebra generated by a chosen basis $\{e_i\}\subset V$, with relations
$e_i\wedge e_j=-e_j\wedge e_i$. 
Denoting multi-indices by roman capital letters $\rI=(i_1,i_2,\ldots,i_k)$, $\rJ$, $\rK$, etc.,
at level $k$, a basis for $V^{\wedge k}$ is given by 
$$e_{\rI}=e_{i_1}\wedge\cdots \wedge e_{i_k}$$
with $\rI$ running through all $k$-subsets
of $\{1, 2,\ldots,n\}$, i.e., $k$-tuples with distinct components. For $M(f)$, $f\in F(S)$, define the matrix 
$$(\spow{M(f)}{k})_{\rI\rJ}=|(M(f)^{\wedge k})_{\rI\rJ}|$$
taking absolute values entry-wise.
It is important to observe that we are not taking the fully symmetric representation of $\End(V)$, 
which would come by looking at the action on boson Fock space, spanned by
symmetric tensors. However, note that the fully symmetric representation is given by maps induced by the
action of $M(f)$ on the algebra generated by commuting variables $\{e_i\}$.
We take this viewpoint as the starting point of the construction of the zeon Fock space, $\Z$, to be
defined presently.

\begin{definition}     
A {\sl zeon algebra\/} is a commutative, associative algebra
generated by elements $e_i$ such that $e_i^2=0$, $i\ge 1$.\\

For a standard zeon algebra, $\Z$, the elements $e_i$ are finite in number, $n$,
and are the basis of an $n$-dimensional vector space, $\V\approx \Q^n$. 
We assume no further relations among the generators $e_i$.
Then the $k^{\rm th}$ {\sl zeon tensor power\/} of $\V$, denoted $\spow{\V}{k}$,
is the degree $k$ component of the graded algebra $\Z$,  with basis 
$$ e_{\rI}=e_{i_1}\cdots e_{i_k}$$
analogously to the exterior power except now the variables commute. The assumptions on the $e_i$ imply that
$\spow{\V}{k}$ is isomorphic to the subspace of symmetric tensors 
spanned by elementary tensors with no repeated factors. {\it As vector spaces},
$$ \spow{\V}{k} \approx \V^{\wedge k} $$

The {\sl zeon Fock space\/} is $\Z$ presented as a graded algebra
$$ {\Z}=\Q\oplus (\bigoplus_{k\ge1}\spow{\V}{k})$$
Since $\V$ is finite-dimensional, $k$ runs from 1 to $n=\dim \V$.
\end{definition}

A linear operator $W \in\End(\V)$ extends to the operator $\spow{W}{k}\in \End(\spow{\V}{k})$.
The {\sl second quantization\/} of $W$ is the induced map on $\Z$. \\

For the exterior algebra, the $\rI\rJ^{\it th}$ component of $W^{\wedge k}$ 
is the determinant of the corresponding submatrix of $W$, with
rows indexed by $\rI$ and columns by $\rJ$. Having dropped the signs,
the $\rI\rJ^{\it th}$ component of $\spow{W}{k}$ is the permanent of the corresponding submatrix of $W$.\\

For $W$ of the form $M(f)$ corresponding to a function $f\in F(S)$, the resulting components of 
$\spow{W}{k}$ are exactly the absolute values of the entries of $W^{\wedge k}$, as we wanted.
At each level $k$, there is an induced map
$$ \End(\V) \to \End(\spow{\V}{k})\ ,\qquad M(f)\to \spow{M(f)}{k}$$
satisfying 

\begin{equation}\label{zhom}
\spow{(M(f_1f_2))}{k}=\spow{(M(f_1)M(f_2))}{k}=\spow{M(f_1)}{k}\spow{M(f_2)}{k}
\end{equation}

giving, for each $k$, a representation of the semigroup $F_\circ(S)$ as endomorphisms
of $\spow{\V}{k}$. However, for general $W_1$, $W_2$, the homomorphism property, \eqref{zhom},
no longer holds, i.e., $ \spow{(W_1W_2)}{k}$ does not necessarily equal $\spow{W_1}{k}\,\spow{W_2}{k}$.
It is not hard to see that a sufficient condition is that
$W_1$ have at most one non-zero entry per column or that $W_2$ have at most one non-zero entry per row. 
For example, if one of them is diagonal,  as well as  the case where both correspond to functions.\\

What is the function, $f_k$, corresponding to $\spow{M(f)}{k}$, i.e., such that $M(f_k)=\spow{M(f)}{k}$ ?
For degree 1, we have from \eqref{Xf}
\begin{equation}\label{xdef}
e_iM(f)=e_{f(i)}
\end{equation}

And for the induced map at degree $k$, taking products in $\Z$,

\begin{eqnarray*}
e_{\rI}\spow{M(f)}{k} &=&(e_{i_1}M(f))\,(e_{i_2}M(f))\cdots (e_{i_k}M(f)) \cr
                                &=& e_{f(i_1)}\,e_{f(i_2)}\cdots e_{f(i_k)}
\end{eqnarray*}

We see that the degree $k$ maps are those induced on $k$-subsets of $S$ mapping
$$\{i_1,\ldots,i_k\} \to \{f(i_1),\ldots,f(i_k)\}$$
with the property that the image in the zeon algebra is zero if $f(i_l)=f(i_m)$ for any pair $i_l,i_m$.
Thus the second quantization of $M(f)$ corresponds to the induced map,
the second quantization of $f$, extending the domain of $f$ from $S$ to
the power set $2^S$.\\

The main features of our construction have been shown. After some preliminaries in the next
subsection, we continue with noncommutative integration, focusing on level 2.

\subsection{The degree 2 component of $\Z$}

Working in degree 2, we denote indices $\rI=(i,j)$, as usual, instead of $(i_1,i_2)$.\\

For given $n$, $X$, $Y$, etc., are vectors in $\spow{\V}{2}\approx\Q^{{n\choose 2}}$.\\

As a vector space, $\V$ is isomorphic to $\Q^n$. Denote by ${\rm Sym}(\V)$ the space of
symmetric matrices acting on $\V$.\\

\begin{definition}     
The mapping  
$${\rm Mat}\colon\,\spow{\V}{2}\to {\rm Sym}(\V)$$
is the linear embedding taking
the vector $X=(x_{ij})$ to the symmetric matrix $\hat X $ with components 
$$\hat X_{ij}=\cases{x_{ij}, & for $i<j$ \cr
                                   0, & for $i=j$\cr}
$$
and the property $\hat X_{ji}=\hat X_{ij}$ fills out the matrix.
\end{definition}

We will use the explicit notation ${\rm Mat}(X)$ as needed for clarity.\\

Equip $\spow{\V}{2}$ with the inner product 
$$\avg X,Y.=\tr \hat X\hat Y\ .$$
Throughout, we use the convention wherein repeated Greek indices are automatically summed over.
So we write
$$\avg X,Y.=X_{\lambda\mu} Y_{\lambda\mu}\ .$$


\section{Basic Identities}

Multiplying $X$ with $u$, we observe that
\begin{equation}\label{eqtrace1}
X\dg u.=\frac12\,\tr \hat X J=u\dg  X.
\end{equation}
Observe also that if $D$ is diagonal, then $\tr D=\tr DJ$.\\

\begin{proposition}\label{propBR}  {\it (Basic Relations)} We have\\

1. $\Mat(X\spow{A}{2}) = \symm{A}{\hat X}-D^+$, where $D^+$ is a diagonal matrix satisfying $\vstrut\tr D^+=\tr \symm{A}{\hat X}$.\\

2. $\Mat(\spow{A}{2}\dg  X.) = \sym{A}{\hat X}-D^-$, where $D^-$ is a diagonal matrix satisfying $\vstrut\tr D^-=\tr \sym{A}{\hat X}$.\\

3. If $A$ and $X$ have nonnegative entries, then $D^+$ and $D^-$ have nonnegative entries. In particular, in that case,
vanishing trace for $D^{\pm}$ implies vanishing of the corresponding matrix.
\end{proposition}

\begin{proof} The components of $X\spow{A}{2}$ are
\begin{eqnarray*}
(X\spow{A}{2})_{\,ij}  &=& \theta_{ij}\theta_{\lambda\mu}
(x_{\lambda\mu}A_{\lambda i}A_{\mu j}+x_{\lambda\mu}A_{\mu i}A_{\lambda j}) \\
 &=& \theta_{ij}(\symm{A}{\hat X})_{\,ij}
\end{eqnarray*}
with the {\sl theta symbol\/} for pairs of single indices
$$\theta_{ij}=\cases{ 1, & if $i<j$\cr  0, & otherwise }$$
Note that the diagonal terms of $\hat X$ vanish anyway.
And $\symm{A}{\hat X}$ will be symmetric if $\hat X$ is.
Since the left-hand side has zero diagonal entries, we can remove the theta symbol and compensate by subtracting off
the diagonal, call it $D^+$. Taking traces yields \#1. And \#2 follows similarly.
\end{proof}

\begin{remark}
Observe that $D^+$ and $D^-$ may be explicitly given by
\begin{eqnarray*}
                    D^+_{\,ii}  &=& 2\,x_{\lambda\mu}A_{\lambda i}A_{\mu i} \\
                    D^-_{\,ii}  &=& 2\,x_{\lambda\mu}A_{i\lambda }A_{i\mu} 
\end{eqnarray*}
where for $D^+$ the $A$ elements are taken within a given column, while for $D^-$, the $A$ elements are in a given row.
\end{remark}\\

\begin{proposition}\label{propXX} Let $X$ and $A$ be nonnegative. Then \\
\begin{eqnarray*}
  \hat X = \symm{A}{\hat X}  &\Rightarrow&  X\spow{A}{2} = X\\
  \hat X = \sym{A}{\hat X}  &\Rightarrow&  \spow{A}{2}\dg  X. = \dg  X.\\
\end{eqnarray*}
\end{proposition}

\begin{proof} We have
$$D^+= \symm{A}{\hat X}-\Mat(X\spow{A}{2})$$
If $\hat X = \symm{A}{\hat X}$, then since $\hat X$ has vanishing trace, $\tr \symm{A}{\hat X}=0$. So
$\tr D^+=0$, hence $D^+=0$, and $\hat X = \symm{A}{\hat X}=\Mat(X\spow{A}{2})$. 
The second implication follows similarly. 
\end{proof}

\section{Trace Identities}

Using equation (\ref{eqtrace1}), we will find some identities for these quantities.\\

\begin{proposition}\label{propTR} We have \\

1. $\displaystyle X\spow{A}{2}\dg  u. = \half \,\tr (\hat X A(J-I)A^*)$ .\\

2. $\displaystyle u\spow{A}{2}\dg  X. = \half \,\tr (\hat X A^*(J-I)A)$ .\\

3. If $A$ is stochastic, then \ $\displaystyle X\spow{A}{2}\dg  u. = \half \,\tr (\hat X (J-AA^*))$.
\end{proposition}

\begin{proof} We have, using equation (\ref{eqtrace1}) and Basic Relation 1,
\begin{eqnarray*}
X\spow{A}{2}\dg  u.  &=& \half \,\tr \Mat(X\spow{A}{2})J\\
 &=& \half\,\tr (\symm{A}{\hat X}J-D^+J) \\
 &=& \half\,\tr (\symm{A}{\hat X}J-\symm{A}{\hat X})
 \end{eqnarray*}
and rearranging terms inside the trace yields \#1. Then \#3 follows since $A$ stochastic implies $AJ=J=JA^*$.
And \#2 follows similarly, using the second Basic Relation in the equation
$\displaystyle u\spow{A}{2}\dg  X. = \half \,\tr \Mat(\spow{A}{2}\dg  X.)J$.
\end{proof}

Using  equation (\ref{eqtrace1}) directly for $X$, we have \\
\begin{eqnarray}
X(I-\spow{A}{2})\dg  u.  &=&  \half \,\tr (\hat X (J-AJA^*+AA^*)) \label{eqbp1}\\
u(I-\spow{A}{2})\dg  X.  &=&  \half \,\tr (\hat X (J-A^*JA+A^*A))\label{eqbp2}
\end{eqnarray}


\subsection{Stochastic case}

For stochastic $A$, equation (\ref{eqbp1}) yields
\begin{lemma} [\rm ``integration-by-parts for zeons"] \label{eqparts}
\begin{equation}
X(I-\spow{A}{2})\dg  u. = \half \,\tr A^*\hat XA
\end{equation}
\end{lemma}

\subsection{Markov-Perron-Frobenius theory}
Let's  recall the basic facts about the convergence of a Markov chain with transition matrix $A$.\\

We denote an invariant distribution for $A$ by $\pi$, i.e. $\pi A=\pi=(p_1,\ldots,p_n)$.\\

$A$ is {\it ergodic\/}, if $A^n$ converges to $\displaystyle \lim_{n\to\infty}A^n = \Omega=\dg  u.\pi$, satisfying $\Omega^2=\Omega=A\Omega=\Omega A$.\\

The {\it state transition diagram\/}, STD,  of $A$ is the directed graph with vertices $\{1,\ldots,n\}$ with
an edge from $i$ to $j$ if $A_{ij}>0$. \\

The notions of irreducibility and aperiodicity in terms of the STD may be taken as definitions.
For an alternative approach with details, see \cite[Ch\th4, \S3]{M}.\\

$A$ is {\it irreducible\/} if for every pair $(i,j)$ there is a  path in the STD with initial vertex $i$ and final vertex $j$.
That is, the STD is strongly connected.\\

In the reducible case, a {\it communicating class\/}, $C$, is a set of states that forms a strongly connected
component of the STD. If there are no transient states, these classes comprise a partition of the set of states.
This is the situation we are considering here.\\

A {\it cycle\/} in the STD is a path with equal initial and final vertices. \\

For $A$ irreducible, it is {\it aperiodic\/} if the greatest common divisor of all cycle lengths is 1.\\

Taking terminology from \cite{S}, say that $A$ is {\it quasi-positive\/} if some power $m\in \N$ of $A$, $A^m$,
has all positive entries. The basic results are these:

\begin{bigthm} {\rm \cite[Ch.\th3, Th.\th2.1]{M},\cite[Th.\th2.9]{S} }$\vstrut$\hfill\break
$A$ is irreducible and aperiodic if and only if it is quasi-positive. 
\end{bigthm}

and

\begin{bigthm} {\rm \cite[Th.\th2.4]{S}, \cite[Ch.\th5, \S2]{D} }$\vstrut$\hfill\break
$A$ is ergodic if and only if it is quasi-positive. 
\end{bigthm}

In other words,\\

{\bf Corollary C\ } \emph{$A$ is ergodic if and only if it is irreducible and aperiodic.} \\

And quasi-positivity serves as a test criterion.\\

For a general setting, see \cite[Ch.\th5, \S2]{D}, where in the Appendix it is remarked that the implication
``quasi-positive implies ergodic" goes back to Markov \cite{MK}.\\

Recall that there may be transient states forming a class, $T$, such that eventually the chain
leaves $T$ and enters a closed ergodic class of states. In this case, $A$ will have a left eigenvector with
eigenvalue 1, i.e., a left-invariant vector, with zero entries. We will not consider this case.\\

Next we derive a new criterion for ergodicity.

\subsection{Ergodicity of a stochastic matrix}

First, a converse result to Proposition \ref{propXX}. 

\begin{proposition}\label{propST} Let  $A$ be stochastic and $X$ nonnegative.\\

1. $X\spow{A}{2} = X \Rightarrow  \hat X = \symm{A}{\hat X}$.\\

2. If $A$ has a strictly positive invariant distribution, $\pi$, then \hfill\break
$\spow{A}{2}\dg  X. = \dg  X.   \Rightarrow  \hat X = \sym{A}{\hat X}\vstrut$.
\end{proposition}

\begin{proof} If $X\spow{A}{2} = X$, then Proposition \ref{propBR}, eq.\th 1, reads
$$\hat X = \symm{A}{\hat X}-D^+$$
Multiply by $\dg  u.$ on the right and by $u$ on the left, using  $A\dg  u.=\dg  u.$ and $uA^*=u$ to get
$$ u\hat X\dg  u. =u\hat X\dg  u.-uD^+\dg  u.$$
So $uD^+\dg  u.=0$ and taking traces yields $\tr(D^+J)=0$, thus $D^+=0$.\\

If $\spow{A}{2}\dg  X. = \dg  X. $, then Proposition \ref{propBR}, eq.\th 2, reads
$$\hat  X = \sym{A}{\hat X}-D^-$$
Multiply by $\dg  \pi.$ on the right and by $\pi$ on the left, using $\pi A=\pi$ and $A^*\dg \pi.=\dg \pi.$ to get
$$ \pi\hat X\dg  \pi. =\pi\hat X\dg  \pi.-\pi D^-\dg  \pi.$$
So $\pi D^-\dg  \pi.=0$ and taking traces yields $\tr(D^-\dg  \pi.\pi)=0$. I.e., $\displaystyle \sum_i D^-_{ii}p_i^2=0$, so
$D^-=0$. 
\end{proof}

With $A$ stochastic, irreducible and aperiodic, $\displaystyle \lim_{n\to\infty}A^n = \Omega=\dg  u.\pi$, satisfying
$\Omega^2=\Omega=A\Omega=\Omega A$. Observe that
\begin{eqnarray*}
\Omega\Omega^* &=& \dg  u.\pi\dg  \pi.u =(\sum p_i^2)\, J\\
\Omega^*\Omega &=& \dg  \pi.u\dg  u.\pi = n\,\dg  \pi.\pi\\
\end{eqnarray*}
both of these having all positive entries.  Thus\\

\begin{proposition}\label{propIRPER} 
Let $A$ be stochastic, irreducible and aperiodic. Then, for nonnegative $X$, either of 
$\hat X=\sym{A}{\hat X}$ or $\hat X=\symm{A}{\hat X}$ implies $X=0$.
\end{proposition}

\begin{proof} If $\hat X=\sym{A}{\hat X}$, then,  inductively for positive integers $k$,
$$ \hat X=A^k\hat X (A^*)^k$$
Letting $k\to\infty$, we have $\hat X=\Omega \hat X \Omega^*$. Taking traces yields 
$\tr \hat X \Omega^*\Omega=0$. Since $\hat X$ is nonnegative and $\Omega^*\Omega$ is a matrix with positive
entries, we have $X=0$. The proof for $\hat X=\symm{A}{\hat X}$ is similar. 
\end{proof}

\begin{theorem}\label{thmIRPER} \hfill\break \nopagebreak

1. Let $A$ have a strictly positive left-invariant vector. Then $\det(I-\spow{A}{2})\ne 0$
implies that $A$ is irreducible and aperiodic. \\

2. If $A$ is irreducible and aperiodic, then $\det(I-\spow{A}{2})\ne 0$.
\end{theorem}

\begin{proof}
For $A$ reducible, let $\{C_0,\ldots,C_{s-1}\}$, $s>1$, denote the communicating classes partitioning the states.
Define $X$ by $x_{ij}=0$ if there is a $k$ such that $i,j\in C_k$, $1$ otherwise, i.e.,
$$ x_{ij}  = 1-\sum_k \chi_{C_k\times C_k}(i,j)$$
We will show that $\spow{A}{2}\dg  X.=\dg  X.$. \\

Step 1. First check that the diagonal vanishes, i.e., $A_{i\lambda}x_{\lambda\mu}A_{i\mu}=0$. By assumption, if $i\in C_k$, then
$\lambda,\mu\in C_k$. But then $x_{\lambda\mu}=0$.\\

Step 2. For $i\ne j$, $(\spow{A}{2}\dg  X.)_{\,ij}=
x_{\lambda\mu}(A_{i\lambda}A_{j\mu}+A_{i\mu}A_{j\lambda})$. Consider the first sum, as the second is similar.
If $i\in C_k$, then the sum is over
$\lambda\in C_k$. If $j\in C_k$, then the sum is over $\mu\in C_k$ as well, $x_{\lambda\mu}=0$, and the result is zero. 
If $j\in C_{k'}$, $k'\ne k$, then $x_{ij}=1$. Writing out the sums, and exchanging indices in the second sum,
since $x_{lm}=1$ as $l$ and $m$ range in different classes,
\begin{eqnarray*}
 \sum_{l\in C_k, m\in C_{k'}\atop l<m} x_{lm}A_{il}A_{jm} &+&
\sum_{l\in C_{k'}, m\in C_{k}\atop l<m} x_{lm}A_{im}A_{jl}\\
&&=\sum_{l\in C_k, m\in C_{k'}\atop l<m} x_{lm}A_{il}A_{jm}+
\sum_{l\in C_{k}, m\in C_{k'}\atop l>m} x_{ml}A_{il}A_{jm}\\
&&=\sum_{l\in C_k, m\in C_{k'}} A_{il}A_{jm}
\end{eqnarray*}

For fixed $m$, the sum over $l\in C_k$ gives all choices that $i$ could map to, 
with $A_{il}=0$ for $l\notin C_k$, so 
$\displaystyle  \sum_{l\in C_k} A_{il}=1$. 
Similarly, summing over $m$ gives 1. Hence, we get $1$ in case $x_{ij}\ne0$ and
$0$ otherwise, as required.\\

If $A$ is irreducible and periodic, with period $p$, say, let $\{C_0,\ldots,C_{p-1}\}$ 
denote the classes partitioning the states such that $i\in C_k$ implies $iA\in C_{k+1}$, with indices modulo $p$.
Define $X$ by $x_{ij}=1$ if there exists $k$ such that $i\in C_k$ and $j\in C_{k+1}$, or $j\in C_k$ and 
$i\in C_{k+1}$. Otherwise, $x_{ij}=0$.  We check that 
$\spow{A}{2}\dg  X.=\dg  X.$.  The steps are similar to that for the case of reducibility:\\

Step 1. First check that the diagonal vanishes, i.e., $A_{i\lambda}x_{\lambda\mu}A_{i\mu}=0$. By periodicity, if $i\in C_{k-1}$, then
$\lambda\in C_k$ and hence $\mu\in C_{k+1}$ or $\mu\in C_{k-1}$, by definition of $X$. But then $A_{i\mu}$ vanishes since $\mu\in C_{k+1}$ requires $i\in C_k$ and $\mu\in C_{k-1}$ requires $i\in C_{k-2}$.\\

Step 2. For $i\ne j$, 
$(\spow{A}{2}\dg  X.)_{\,ij}=
x_{\lambda\mu}(A_{i\lambda}A_{j\mu}+A_{i\mu}A_{j\lambda})$.
Let $i\in C_k$. Then $\lambda\in C_{k+1}$ gives $\mu\in C_k$ or $\mu\in C_{k+2}$, finally,
$j\in C_{k-1}$ or $j\in C_{k+1}$. And then $x_{ij}=1$. The other case is $\mu\in C_{k+1}$ which gives
$\lambda\in C_{k}$ or $\lambda\in C_{k+2}$ and again $j\in C_{k-1}$ or $j\in C_{k+1}$, with
the conclusion $x_{ij}=1$. Otherwise we get $x_{ij}=0$.
Writing out the sums, in all cases, $x_{lm}=1$. We get, for $j\in C_{k-1}$,
$$ \sum_{l\in C_{k+1}, m\in C_{k}} A_{il}A_{jm}
+ \sum_{l\in C_{k}, m\in C_{k+1}} A_{im}A_{jl} $$
and for  $j\in C_{k+1}$,
$$ \sum_{l\in C_{k+1}, m\in C_{k+2}} A_{il}A_{jm}
+ \sum_{l\in C_{k+2}, m\in C_{k+1}} A_{im}A_{jl} $$
Note that the four sums have no overlapping terms, except if $p=2$, then both cases for $j$ are the same.
In the first sum, for fixed $m$, the sum over $l$ gives all choices that $i$ could map to, yielding
1, and summing over $m$ gives 1 as well. The other three sums are similar, adding in each case to 1.
Hence, we get $1$ in case $x_{ij}\ne0$ and $0$ otherwise, as required.\\

Suppose $A$ is aperiodic and irreducible.  
By Perron-Frobenius applied to $\spow{A}{2}$,
its top nonnegative eigenvalue, $\lambda_0$, has a nonnegative eigenvector, $X$, \cite[Ch.\th 1, Th.\th4.4]{M}.
First we show that $\lambda_0\le1$, so that if $1$ is an eigenvalue, it is the top nonnegative eigenvalue.
By Lemma \ref{eqparts}, we have, with $X\dg u.>0$,
$$\displaystyle X(I-\spow{A}{2})\dg  u. = X(1-\lambda_0)\dg u. = 
(1-\lambda_0)X\dg u. = \half \,\tr (A^*\hat X A)\ge 0\ .$$
Now, if $X\spow{A}{2} =X$, then $\lambda_0=1$. So the corresponding eigenvector $X$ is
nonnegative. Proposition \ref{propST} says that $X\spow{A}{2} =X$ implies $\hat X=\symm{A}{\hat X}$. 
Then Proposition \ref{propIRPER} yields $X=0$. 
\end{proof}

\subsection{Some examples}

Here are some examples illustrating a variety of cases.\\

\begin{example}
Let $$A=\left[ \begin {array}{ccc} 1/4&1/4&1/2\\\noalign{\medskip}1/4&1/4&1/2
\\\noalign{\medskip}0&0&1\end {array} \right]$$
then
$$\spow{A}{2}=\left[ \begin {array}{ccc} 1/8&1/4&1/4\\\noalign{\medskip}0&1/4&1/4
\\\noalign{\medskip}0&1/4&1/4\end {array} \right]$$
with $\det(I-\spow{A}{2})=7/16\ne0$. The limit
$$ \lim_{n\to\infty} A^n= \left[\matrix{0&0&1\cr 0&0&1\cr 0&0&1\cr}\right]$$
exists, with zeros corresponding to the transient states $\{1,2\}$.
\end{example}

\begin{example}     
Here there are two absorbing states, $\{2,4\}$, with transient $\{3\}$ having the possibility to go to either of them.\\
$$A=\left[ \begin {array}{cccc} 1/2&1/2&0&0\\\noalign{\medskip}0&1&0&0
\\\noalign{\medskip}0&1/2&0&1/2\\\noalign{\medskip}0&0&0&1\end {array}
 \right] $$
then
$$\spow{A}{2}= \left[ \begin {array}{cccccc} 1/2&0&0&0&0&0\\\noalign{\medskip}1/4&0&
1/4&0&1/4&0\\\noalign{\medskip}0&0&1/2&0&1/2&0\\\noalign{\medskip}0&0&0
&0&1/2&0\\\noalign{\medskip}0&0&0&0&1&0\\\noalign{\medskip}0&0&0&0&1/2
&0\end {array} \right]$$
with $\det(I-\spow{A}{2})=0$. The limit
$$ \lim_{n\to\infty} A^n= \left[ \begin {array}{cccc} 0&1&0&0\\\noalign{\medskip}0&1&0&0
\\\noalign{\medskip}0&1/2&0&1/2\\\noalign{\medskip}0&0&0&1\end {array}
 \right]$$
exists. A left-invariant vector of $A$ is a linear combination of
$$\{[\,0,1,0,0\,],\,[\,0,0,0,1\,]\}$$
while a left-invariant vector of $\spow{A}{2}$ is a multiple of
$$[\,0, 0, 0,0, 1, 0\,]$$
having a 1 in the $(2,4)$ spot.
A right-invariant vector of $\spow{A}{2}$ is a multiple of
$$\left[\matrix{0&1&2&1&2&1}\right]^\dagger$$
noting that $\spow{A}{2}$ does not necessarily have constant row sums.
\end{example}

\begin{example}     
Here's a reducible case, no transients.
$$A=\left[ \begin {array}{ccccc} 1/2&1/2&0&0&0\\\noalign{\medskip}1/2&1/2
&0&0&0\\\noalign{\medskip}0&0&0&1/2&1/2\\\noalign{\medskip}0&0&1/2&0&1
/2\\\noalign{\medskip}0&0&1/2&1/2&0\end {array} \right]$$
A left-invariant vector of $A$ has the form
$$[\,x_2,x_2,x_1,x_1,x_1\,] $$
with $x_i$ arbitrary. Closed classes are $\{1,2\},\{3,4,5\}$.
Solutions $X$ to $\spow{A}{2}\dg X.=\dg X.$ satisfy
$$\hat X=  \left[ \begin {array}{ccccc} 0&0&w_{{1}}&w_{{1}}&w_{{1}}
\\\noalign{\medskip}0&0&w_{{1}}&w_{{1}}&w_{{1}}\\\noalign{\medskip}w_{
{1}}&w_{{1}}&0&0&0\\\noalign{\medskip}w_{{1}}&w_{{1}}&0&0&0
\\\noalign{\medskip}w_{{1}}&w_{{1}}&0&0&0\end {array} \right]$$
with arbitrary parameters $w_i$. This is the solution appearing in the above proof.
\end{example}

\begin{example}     
This case is periodic, without transients.
$$A=\left[ \begin {array}{ccccc} 0&1&0&0&0\\\noalign{\medskip}0&0&1/2&1/2
&0\\\noalign{\medskip}0&0&0&0&1\\\noalign{\medskip}0&0&0&0&1
\\\noalign{\medskip}1&0&0&0&0\end {array} \right] $$
Any left-invariant vector of $A$ is a multiple of
$$[\,2, 2, 1, 1, 2\,] $$
Solutions $X$ to $\spow{A}{2}\dg X.=\dg X.$ satisfy
$$\hat X= \left[ \begin {array}{ccccc} 0&w_{{2}}&w_{{1}}&w_{{1}}&w_{{2}}
\\\noalign{\medskip}w_{{2}}&0&w_{{2}}&w_{{2}}&w_{{1}}
\\\noalign{\medskip}w_{{1}}&w_{{2}}&0&0&w_{{2}}\\\noalign{\medskip}w_{
{1}}&w_{{2}}&0&0&w_{{2}}\\\noalign{\medskip}w_{{2}}&w_{{1}}&w_{{2}}&w_
{{2}}&0\end {array} \right]$$
with arbitrary $w_i$. The periodic classes are  $\{1\},\{2\},\{3,4\},\{5\}$. For a fixed $\delta$,
a basic right-invariant vector for $\spow{A}{2}$ is given by $x_{ij}=1$ if ${\rm dist\,}(i,j)=\delta$, $0$
otherwise. Here, distance is the shortest value $|k-k'|$, labelling the classes consecutively modulo 4,
with $i\in C_k$, $j\in C_{k'}$. The analogous construction works for general period $p$. In the above proof,
we specialized to $\delta=1$.
\end{example}

\begin{example}     
Our concluding example is both reducible and periodic. 
$$A=\left[ \begin {array}{cccccc} 0&1&0&0&0&0\\\noalign{\medskip}0&0&0&0&
1&0\\\noalign{\medskip}0&0&0&1&0&0\\\noalign{\medskip}0&0&1&0&0&0
\\\noalign{\medskip}0&1/2&0&0&0&1/2\\\noalign{\medskip}1&0&0&0&0&0
\end {array} \right] $$
with left-invariant vectors of the form
$$[\,x_{{2}},2\,x_{{2}},x_{{1}},x_{{1}},2\,x_{{2}},x_{{2}}\,]$$
for arbitrary $x_i$. The periodic classes are  $\{1,5\},\{2,6\},\{3,4\}$, with closed classes
$\{1,2,5,6\},\{3,4\}$.
Solutions $X$ to $\spow{A}{2}\dg X.=\dg X.$ have the form
$$\hat X=  \left[ \begin {array}{cccccc} 0&w_{{4}}&w_{{2}}&w_{{3}}&0&w_{{4}}
\\\noalign{\medskip}w_{{4}}&0&w_{{3}}&w_{{2}}&w_{{4}}&0
\\\noalign{\medskip}w_{{2}}&w_{{3}}&0&w_{{1}}&w_{{2}}&w_{{3}}
\\\noalign{\medskip}w_{{3}}&w_{{2}}&w_{{1}}&0&w_{{3}}&w_{{2}}
\\\noalign{\medskip}0&w_{{4}}&w_{{2}}&w_{{3}}&0&w_{{4}}
\\\noalign{\medskip}w_{{4}}&0&w_{{3}}&w_{{2}}&w_{{4}}&0\end {array}
 \right]$$
with the $w_i$ arbitrary.
\end{example}

\section{Conclusion}
After introducing the zeon Fock space, we have found some basic identities and properties for the corresponding
noncommutative integration at level 2. These are used to find a criterion for a Markov chain to 
be irreducible and aperiodic.\\

There are interesting applications to semigroups of matrices representing functions acting on a finite set.
These can be related to directed graphs of constant out-degree. This work will appear separately.\\

Extensions to countable state space look to be interesting possibilities for further work,
\cite[Ch.\th 4 \S4]{D}.\\

\begin{acknowledgment}
The discussions and insight of G.\th Budzban are deeply appreciated for making this work possible.
We thank J.\th Kocik for useful consultations.
\end{acknowledgment}

\end{document}